\documentclass[twocolumn,showpacs,preprintnumbers,amsmath,amssymb,superscriptaddress,floatfix]{revtex4}

\usepackage{graphicx} 
\usepackage{dcolumn} 
\usepackage{bm} 
\usepackage{amsmath}

\begin{document}

\preprint{}

\title{Charge pumping in carbon nanotube quantum dots}

\author{M.R.~Buitelaar}
\affiliation{Cavendish Laboratory, University of Cambridge,
Cambridge, CB3 0HE, UK}
\author{V.~Kashcheyevs}
\affiliation{Institute for Solid State Physics, University of
Latvia, Riga, LV-1063, Latvia} \affiliation{Faculty of Physics and
Mathematics, University of Latvia, Riga, LV-1002, Latvia}
\author{P.J.~Leek}
\author{V.I.~Talyanskii}
\author{C.G.~Smith}
\author{D.~Anderson}
\author{G.A.C.~Jones}
\affiliation{Cavendish Laboratory, University of Cambridge,
Cambridge, CB3 0HE, UK}
\author{J.~Wei}
\author{D.H.~Cobden}
\affiliation{Department of Physics, University of Washington,
Seattle, Washington 98195-1560, USA}

\date{\today}

\begin{abstract}
We investigate charge pumping in carbon nanotube quantum dots
driven by the electric field of a surface acoustic wave. We find
that at small driving amplitudes, the pumped current reverses
polarity as the conductance is tuned through a Coulomb blockade
peak using a gate electrode. We study the behavior as a function
of wave amplitude, frequency and direction and develop a model in
which our results can be understood as resulting from adiabatic
charge redistribution between the leads and quantum dots on the
nanotube.
\end{abstract}

\pacs{73.63.Kv, 73.23.Hk, 73.63.Fg, 85.35.Kt, 72.50.+b}

\maketitle


The dynamics of charge transport at the level of single electrons
or electron spins in nanoscale systems such as quantum dots is
important from both applied and fundamental viewpoints. A
sensitive probe of the response of a quantum dot to changing
external parameters (e.g. a gate voltage or magnetic field) is the
dc charge pumping current generated in the absence of an applied
bias \cite{Thouless}. A general result for open, non-interacting
electron systems is the Brouwer formula which relates the pumped
current to the scattering matrix of the system \cite{Brouwer}. In
interacting systems, correlations add complexity and modify the
predictions. Recent theoretical work has considered pumping in
quantum dots with weak interactions \cite{Aleiner,Brouwer2}, in
the Kondo regime \cite{Aono}, in the Coulomb blockade regime
\cite{Splettstoesser1}, and for superconducting leads
\cite{Blaauboer,Splettstoesser3}. For the case of interacting
electrons in one dimension (the Luttinger liquid), current
quantization \cite{Talyanskii,Novikov} and pure spin currents
\cite{Citro} have been predicted.

Experimentally, the rich variety of predicted phenomena has
remained largely unexplored \cite{Pothier, Switkes}. A promising
system to test the various theories of charge pumping is carbon
nanotubes, since all of the transport regimes mentioned above have
already been realized in conventional nanotube devices
\cite{Nygard,Bockrath,Buitelaar2}. In this Letter we pursue the
potential of nanotubes, demonstrating charge pumping in nanotube
quantum dots up to frequencies of 2.6 GHz and present a
theoretical description of the observed effects.

While charge pumping could in principle be realized by modulating
side or top gates that are capacitively coupled to the nanotubes
\cite{Wei}, we adopt a different approach in which a surface
acoustic wave (SAW) pumps charge through a carbon nanotube
\cite{Leek, Buitelaar, Ebbecke, Shin}. An individual nanotube is
contacted on a piezoelectric quartz substrate by palladium source
and drain electrode that are separated by 5 $\mu$m (Fig.~1(c)
lower right inset). A side gate electrode is used to vary the
electrostatic potential in the nanotube. Several mm beyond each
contact are two SAW transducers (Fig.~1(c) left inset) having
resonant frequencies $f_{\text{SAW}}$ of about 2.6 GHz and 544
MHz. The electric field accompanying the SAW pumps charge through
the nanotube. While the use of SAWs restricts experiments to the
transducers' resonant frequencies, it avoids direct capacitive
coupling between the high-frequency and device electrodes and
enables a clear distinction between the signal due to the SAW
(possible only at the transducers' resonant frequencies) and
rectified currents from radiated fields (possible at all
frequencies). Details of sample fabrication, transducer operation
and propagating SAW fields were described in
Ref.~\cite{Buitelaar}.

Fig.~1(a) shows the dc transport properties of the nanotube at
temperature $T = 5$ K. The approximate periodicity and large
charging energy $U_C \sim 10-15$ meV observed indicate that the
nanotube is divided into two (or more) sections such that the
conductance is dominated by Coulomb blockade of a single electron
puddle that is much smaller than the 5 $\mu$m source-drain
separation. This also follows from the suppression of the
linear-response conductance and additional features (arrows)
observed in the differential conductance (see inset to Fig.~1(a))
\cite{Park}. Fig.~1(c) shows the induced current,
$I_{\text{SAW}}$, in the presence of a SAW field at 2607 MHz in
the absence of a source-drain bias voltage. For a SAW velocity
$v_{\text{SAW}}\approx 3200$ m/s on quartz this corresponds to a
SAW wavelength $\lambda_{\text{SAW}} =
v_{\text{SAW}}/f_{\text{SAW}} \sim$ 1.2 $\mu$m \cite{Buitelaar}.
When a low power $P_{\text{SAW}}$ is applied to the SAW
transducers, a dc current is induced whose direction alternates as
a sensitive function of gate voltage $V_g$. The peak-and-dip
features in the current are clearly correlated with the Coulomb
blockade peaks, and the current changes polarity on reversal of
the SAW direction.

\begin{figure}
\includegraphics[width=80mm]{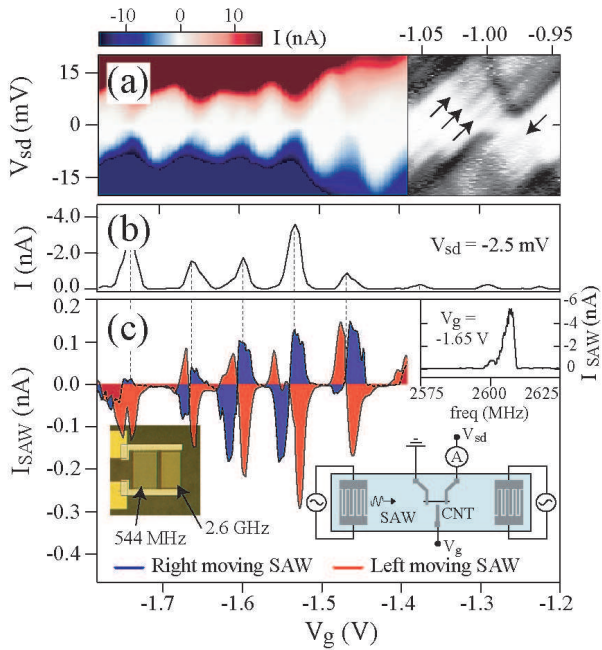}
\caption{\label{Fig1} (a) Color scale representation of the
current as a function of $V_{sd}$ and $V_g$. Inset: differential
conductance around a Coulomb peak (dark = more conductive) (b)
Current trace at $V_{sd} = -2.5$ mV showing Coulomb blockade
oscillations. (c) SAW induced current as a function of $V_g$ in
the absence of an applied bias. In red is the current produced by
the left-moving SAW at 2602 MHz with $P_{\text{SAW}} = -15$ dBm.
In blue is the current produced by the right-moving SAW at 2607
MHz with $P_{\text{SAW}} = -10$ dBm. Right inset: simplified
schematic of the device. Left inset: photograph of one of the
transducers, which is able to generate SAWs at frequencies of both
2607 MHz and 544 MHz. Top inset: frequency dependence of the
current at $V_g = -1.65$ V and $P_{\text{SAW}} = 15$ dBm.}
\end{figure}

\begin{figure}
\includegraphics[width=75mm]{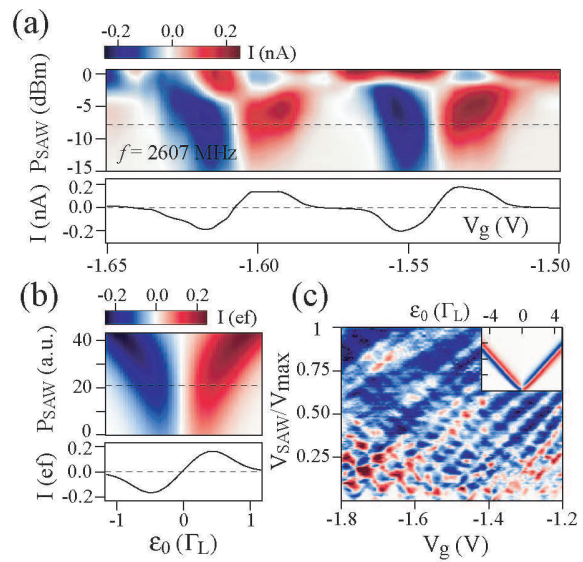}
\caption{\label{Fig2} (a) Color scale representation of the
SAW-induced current as function of $V_{sd}$ and $V_g$, showing
sign reversal as $V_g$ crosses a Coulomb blockade peak. The line
trace is taken at $P_{\text{SAW}} = -8$ dBm. (b) Calculated
current using the model explained in the main text. (c) Derivative
of the SAW induced current with respect to the SAW amplitude as a
function of $V_g$ and (normalized) $V_{\text{SAW}}$. The maximum
SAW amplitude is estimated to be $\sim 100-150$ mV. Inset:
$dI/dV_{\text{SAW}}$ as calculated in the model, showing the same
linear peak splitting.}
\end{figure}

These features were studied in more detail as a function of SAW
amplitude and for both available SAW frequencies. Figure 2(a)
shows $I_{\text{SAW}}$ as a function of $P_{\text{SAW}}$. The
peak-and-dip features increase in magnitude and move outwards as
$P_{\text{SAW}}$ is increased. When we plot the derivative of the
current with respect to the SAW amplitude $V_{\text{SAW}} \propto
P_{\text{SAW}}^{1/2}$, as in Fig.~2(c), it becomes apparent that
the dominant features move essentially linearly with
$V_{\text{SAW}}$. At the highest applied SAW powers (up to 20 dBm
or 100 mW) the features merge into a background of current in the
direction of the SAW across the whole gate voltage range
\cite{Leek}.

Assuming that the dominant action of the SAW is to modulate the
electrostatic potential at the nanotube, we can estimate the
actual value of $V_{\text{SAW}}$ from the linear evolution of the
current features with amplitude and the gate efficiency deduced
from the dc Coulomb blockade pattern. We find a SAW amplitude
$V_{\text{SAW}} \simeq 150$ mV at the maximum power
$P_{\text{SAW}} = 20$ dBm applied to the left transducer. This
agrees with estimates from transducer impedance calculations and
direct transducer transmittance measurements \cite{Buitelaar}. The
inset to Fig.~1(c) shows $I_{\text{SAW}}$ as a function of
frequency. A peak is observed at the transducer resonant frequency
only, demonstrating that the observed current features do not
result from directly radiated fields or from photon-assisted
tunneling (see also Refs.~\cite{Leek,Buitelaar}).

All the observed behavior can be explained by a model in which
disorder in the nanotube causes it to contain two or more
localized electron puddles in series. The SAW modulates the
energies of the states in these puddles and adiabatic charge
redistribution between the puddles and the leads gives rise to a
dc current. We first discuss a minimal model in which there are
two puddles, each acting as a quantum dot with a single spinless
level (see Fig.~3(a)), before introducing asymmetries in the
tunnel couplings and multiple levels in the dots to better reflect
the likely experimental situation. Such a simple two-level system
provides a useful approximation to the Coulomb blockade physics of
tunnel-coupled double dots; see e.g Ref.~\cite{Wiel}. The only
parameters that enter the model are the tunnel couplings
$\Gamma_L$ and $\Gamma_R$ of the left and right dots to left and
right leads respectively, $V_{\text{SAW}}$, the phase difference
$\phi$ of the SAW between the two dots, and the tunnel coupling
$\Delta$ between them. The instantaneous effective Hamiltonian of
this double-dot system can be written as
\begin{align}
\mathcal{H}_d = \begin{bmatrix}
                  \varepsilon_1-i \Gamma_L /2 & \Delta/2 \\
                  \Delta/2 &  \varepsilon_2 - i \Gamma_R /2 \\
                \end{bmatrix} \, .
\end{align}
For any periodic time dependence of the dot energies
$\varepsilon_1$ and $\varepsilon_2$, the adiabatic current
\cite{Buttiker,Brouwer,Entin,Kashcheyevs} from left to right can
be calculated as  a surface integral \cite{Brouwer}, $I = e f
\int_A R(\varepsilon_1, \varepsilon_2)~d \varepsilon_1 d
\varepsilon_2$, over the area  $A$ enclosed parametrically by the
pumping trajectory in the $(\varepsilon_1, \varepsilon_2)$-plane.

The response function $R$ can be obtained from Eq.~1, see e.g.
Ref.~\cite{Kashcheyevs}, and in the zero-temperature limit takes a
straightforward form, $R(\varepsilon_1, \varepsilon_2)=$
\begin{equation}
\nonumber \frac{-(32/\pi) \Delta^2 \Gamma_L \Gamma_R
\bigl(\varepsilon_1 \Gamma_R +\varepsilon_2 \Gamma_L
\bigr)}{\bigl[2 \Gamma_L \Gamma_R \Delta^2+4 \varepsilon_1^2
\Gamma_R^2 +4 \varepsilon_2^2 \Gamma_L^2
    + \bigl(\Delta^2-4 \varepsilon_1
   \varepsilon_2\bigr){}^2+\Gamma_L^2 \Gamma_R^2 \bigr]^2
},
\end{equation}
The response function is shown graphically in Fig.~3(b,c) for
different parameter values. It exhibits one pronounced minimum
(blue) and one maximum (red). For the pumping trajectory, we
assume the SAW periodically modulates the levels according to
\begin{equation}\label{eq:trajectory}
\begin{split}
\varepsilon_1 & = - \delta/2 + \alpha_1 e V_g  + eV_{\text{SAW}} \cos( 2 \pi f t) \, , \\
\varepsilon_2  &= + \delta/2 + \alpha_2 e V_g + eV_{\text{SAW}}
\cos( 2 \pi f t + \phi) \, .
\end{split}
\end{equation}
Here $\alpha_1$ and $\alpha_2$ are the coupling efficiencies of
the gate to the two dots and $\delta$ is a level offset parameter.
The effect of the SAW on the tunnel barriers is thought to be
negligible for the small SAW amplitudes considered here. For the
case of symmetric coupling $\Gamma_L = \Gamma_R$ (Fig.~3(b)), we
see that the peaks of negative and positive pumping current, as
seen in Fig.~2(b) correspond to elliptical trajectories around the
two triple points in the stability diagram of the double dot
\cite{Wiel,Naber}. In obvious notation we may represent these as
the $(0,0)\rightarrow(0,1)\rightarrow(1,0)\rightarrow(0,0)$
electron and $(1,1)\rightarrow
(1,0)\rightarrow(0,1)\rightarrow(1,1)$ hole cycles, respectively.

One property of the pumping current that follows immediately is
that it changes polarity when the direction of the pumping contour
reverses, i.e., when the SAW direction is reversed.  Another is
that for small SAW amplitudes, $I \propto V_{\text{SAW}}^2 \sin
\phi$ \cite{Brouwer}, because in this limit
$R(\varepsilon_1,\varepsilon_2)$ is approximately constant within
the area $A= \pi e^2 V_{\text{SAW}}^2 \sin \phi$ enclosed by the
trajectories.

\begin{figure}
\includegraphics[width=73mm]{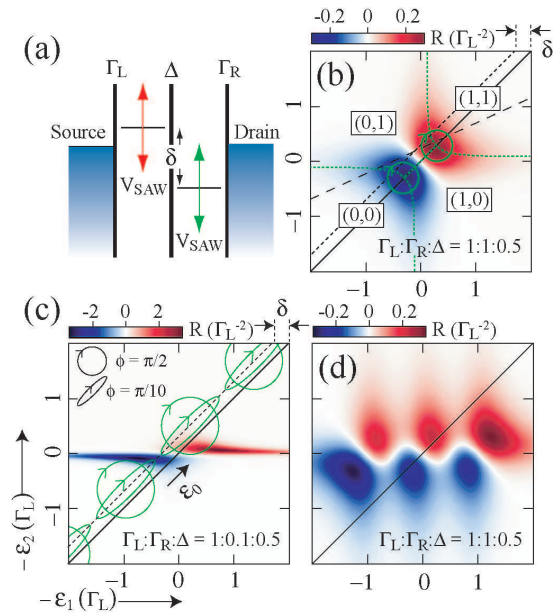}
\caption{\label{Fig3} (a) Schematic of a double dot in which the
single-electron states are periodically modulated by the SAW with
phase difference $\phi$. (b) Color scale representation of the
function $R(\varepsilon_1,\varepsilon_2)$ for symmetric coupling
to the leads. The SAW-induced current is proportional to the
integral of this function over the area traversed in
$[\varepsilon_1,\varepsilon_2]$ space. For $\phi = \pi /2$ the
trajectories are circles with diameters proportional to
$V_{\text{SAW}}$; for $\phi = \pi /10$ they are narrow ellipses.
The ordered pairs ($n$,$m$) indicate the electron occupancy of the
quantum dots. (c) Same as panel (b) for asymmetric coupling to the
leads. The direction $\varepsilon_0$ corresponds to the case of
equal coupling of the dots to the gate, $\alpha_1 = \alpha_2$. (d)
Calculation of $R(\varepsilon_1,\varepsilon_2)$ with 3 levels on
the left dot taken to have spacing equal to $2 \Delta$ and
identical couplings to the level in the other dot.}
\end{figure}

As $V_g$ is varied, the center of the pumping trajectory follows a
straight line which for the simple case $\alpha_1 = \alpha_2$ and
$\delta = 0$ is the solid diagonal in Fig.~3(b)-(d). In this case,
the current traces $I(V_g)$ exhibit symmetric sign-reversing
features. If the dots are single-level and completely symmetric,
i.e., $\Gamma_L = \Gamma_R$ and $\delta =0$, then the current
traces for different $P_{\text{SAW}}$ are dependent only on the
ratio $\Delta/\Gamma$. This simple situation already matches the
experimental data quite well, as shown in Fig.~2(b), taking
$\Delta/\Gamma =0.5$.  Figure 3 also illustrates the effects of
asymmetries. A nonzero level offset $\delta$ shifts the diagonal
line [see e.g. the dotted line in Fig.~3(b)], while asymmetric
coupling to the gate, $\alpha_1 \neq \alpha_2$, changes the slope
(dashed line). The result is an asymmetry between the positive and
negative current peaks which is most pronounced in the weak
pumping regime. For larger SAW powers, such that $eV_{\text{SAW}}
\geq \delta$, the difference between positive and negative peaks
begins to average out. In the limit of strong pumping, the pumping
trajectories encompass both positive and negative parts of
$R(\varepsilon_1, \varepsilon_2)$ whose contributions to the
pumping tend to cancel. This explains the experimental observation
(Fig.~2(c)) that the opposite-sign peak pairs in the pumping
current vs $V_g$ move apart linearly in $V_{\text{SAW}}$: the peak
current occurs where the integral of $R$ is maximal as a function
of $\varepsilon_0$, and this occurs at $|\varepsilon_0| \sim
eV_{\text{SAW}}$

The qualitative behavior of the model is quite insensitive to
details such as asymmetry in the tunnel couplings or a
multiplicity of levels in the dots. In particular the experimental
situation in which the conductance is dominated by a single
quantum dot can be modelled by taking a single level on the right
dot and multiple levels on the left, as shown in Fig.~3(d). The
sign-reversing nature of the $I_{\text{SAW}}$-$V_g$ traces is
still a robust feature although signatures of extra levels may
appear for certain pumping trajectories \cite{levels}.

\begin{figure}
\includegraphics[width=75mm]{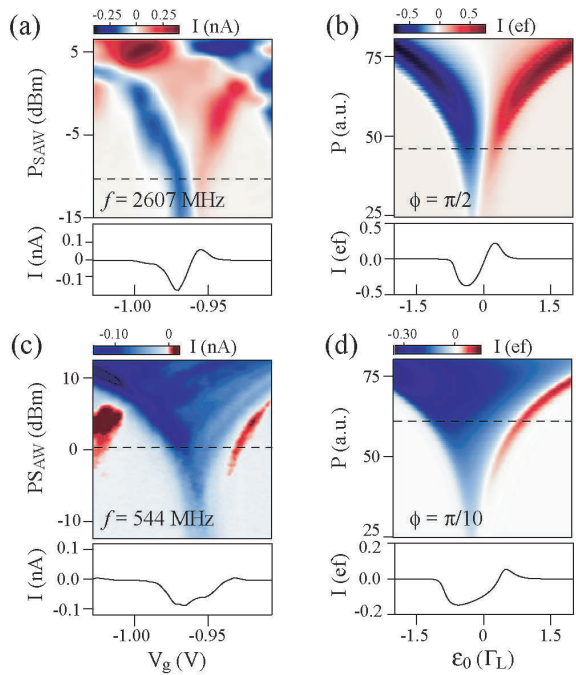}
\caption{\label{Fig4} (a) Color scale representation of the
SAW-induced current as a function of SAW power and $V_g$ for
$f_{\text{SAW}} = 2607$ MHz. (b) Calculated SAW current for a
double QD model with an asymmetry between the tunnel couplings
$\Gamma_L$ and $\Gamma_R$, as discussed in the text. The
parameters are $\Gamma_L : \Gamma_R : \Delta : \delta = 1.0 : 0.1
: 0.5 : 0.3$, and the phase difference $\phi = \pi /2$. (c)
SAW-induced current for the same resonance peak as in panel (a)
for $f_{\text{SAW}} = 544$ MHz. (d) Calculated SAW current as in
panel (b) but for a phase difference $\phi = \pi /10$
\cite{phase}}
\end{figure}

Finally, we examine whether the experimental effects of changing
$\lambda_{\text{SAW}}$ are consistent with the model. Fig. 4(a)
shows $I_{\text{SAW}}$ at $f_{\text{SAW}} = 2607$ MHz in the
vicinity of a Coulomb peak. A slight asymmetry between positive
and negative peaks is seen for small $P_{\text{SAW}}$. In the
model this is reproduced (see Fig. 4(b)) by, for instance,
assuming an asymmetric coupling to the leads and a small offset
$\delta$, corresponding to the situation depicted in Fig.~3(c).
Other kinds of asymmetry such as $\alpha_1 \neq \alpha_2$ have a
similar result. Fig. 4(c) shows the current at $f_{\text{SAW}} =
544$ MHz, using the second pair of transducers with larger finger
spacing. The current is now nearly always of negative polarity.
Fig.~4(d) shows the corresponding model prediction, which agrees
very well. Here, the effect of decreasing $f_{\text{SAW}}$ is just
to reduce the phase difference between the dots ($\phi \rightarrow
0$ when $\lambda_{\text{SAW}}$ becomes much larger than the device
dimensions). This reduces the width and area of the elliptical
pumping trajectory. The result can be understood qualitatively
from Fig.~3(c) noting that the level offset $\delta$ shifts the
line followed by the ellipse center towards the negative (blue)
regions of $R(\varepsilon_1, \varepsilon_2)$. For large SAW powers
and circular pumping trajectories, the enclosed area includes a
domain where $R(\varepsilon_1, \varepsilon_2)$ is positive.
However, after decreasing $P_{\text{SAW}}$ (smaller radius) or
$f_{\text{SAW}}$ (narrower ellipse) the integral only includes
regions of negative $R(\varepsilon_1, \varepsilon_2)$, enhancing
the asymmetry. Note that the enhancement of the asymmetry in
$I_{\text{SAW}}$ when decreasing SAW power or frequency is
expected whenever the center of the pumping trajectory follows a
line in $[\varepsilon_1, \varepsilon_2]$ parameter space that is
asymmetric with respect to the minima and maxima of
$R(\varepsilon_1, \varepsilon_2)$, irrespective of the precise
form of this function. The results in Fig.~4 are therefore
consistent with the model also when in the experiment
$R(\varepsilon_1, \varepsilon_2)$ is significantly more complex
than the model situations shown in Fig.~3. To completely describe
the experiment (the model actually underestimates the amount of
pumped current and does not account for the fine structure
observed at 544 MHz) would require a better knowledge of the
microscopic details of the device than we have at present.

We thank A. Aharony, O. Entin-Wohlman, and L. Levitov for
discussions. MB has been supported by the UK QIP IRC
(GR/S82176/01). VK has been supported by Latvian Council of
Science, European Social Fund and German-Israeli Project
Cooperation (DIP).



\begin{thebibliography}{99}
\bibitem{levels}
This also holds when the level spacing on the left dot is small
compared to the temperature or $\Gamma_L$ and the features in
$R(\varepsilon_1, \varepsilon_2)$ merge into two horizontal bands
of opposite polarity.
\bibitem{phase}
In the model we assume a reduction of $\phi$ by a factor of five,
corresponding to the fivefold increase of $\lambda_{\text{SAW}}$.
The correspondence between $\phi$ and $\lambda_{\text{SAW}}$ is
uniquely defined only when $\lambda_{\text{SAW}}$ is larger than
twice the distance between the centers of the two dots.
\end{thebibliography}
\end{document}